\documentclass[aps,prl,notitlepage,10pt,onecolumn,superscriptaddress,nofootinbib,floatfix]{revtex4-2}
\pdfoutput=1
\usepackage[utf8]{inputenc}
\usepackage[T1]{fontenc}
\usepackage[english]{babel}
\usepackage{graphicx} 
\usepackage{xcolor}
\usepackage[bookmarks=false,colorlinks=true]{hyperref}

\begin{document}

\title{Metrology of Gravitational Effects with Mechanical Quantum Systems}
\author{Daniel Braun}
\affiliation{Institute for Theoretical Physics, Eberhard Karls Universität Tübingen, Auf der Morgenstelle 14, 72076 Tübingen}

\author{Marta M.~Marchese}
\affiliation{Naturwissenschaftlich-Technische Fakultät, Universität Siegen, Walter-Flex-Str.~3, 57068 Siegen, Germany}

\author{Stefan Nimmrichter}
\affiliation{Naturwissenschaftlich-Technische Fakultät, Universität Siegen, Walter-Flex-Str.~3, 57068 Siegen, Germany}

\author{Sofia Qvarfort}
\affiliation{Nordita, KTH Royal Institute of Technology and Stockholm University, Hannes Alfv\'{e}ns v\"{a}g 12, SE-106 91 Stockholm, Sweden}
\affiliation{Department of Physics, Stockholm University, AlbaNova University Center, SE-106 91 Stockholm, Sweden}

\author{Dennis R\"atzel}
\affiliation{ZARM, Unversität Bremen, Am Fallturm 2, 28359 Bremen, Germany}

\author{Hendrik Ulbricht}
\affiliation{School of Physics and Astronomy, University of Southampton, Southampton SO17 1BJ, United Kingdom}

\date{January 2025}

\maketitle

\section{Status}

Here we discuss how massive mechanical quantum systems such as mechanical resonators where the center of mass motion is coupled to optical, electric or magnetic fields can be used for metrology, and especially to measure gravity. This may include suspended mechanical systems {and} levitated mechanical systems. {Systems may also go through a phase of free fall during the experimental protocol.}
{The common advantage of systems considered here is that they carry significant mass which makes them sensitive to effects of gravity.}
Furthermore, 
their center of mass motion can be controlled {or read out} at the quantum level, allowing for application of quantum metrology. Gravitational wave detection at LIGO is the most prominent example {for readout of a mechanical system at the quantum level}, harnessing theoretical methods for quantum parameter estimation and state estimation \cite{aasi2015advanced}. Extensive reviews of such systems and their use to detect gravity can be found elsewhere \cite{aspelmeyer2014cavity,bose2023massive}. This article expresses the view of the authors on the topic and is not intended to be complete.

The interest of the quantum optics community in mechanical quantum systems for metrological applications has increased strongly over the last two decades, as shown by the growing number of proposals \cite{bose2023massive}. This development is caused by the great achievements of experimentalists in bringing larger and larger masses into the quantum regime including free-falling systems like 
molecules \cite{hornberger2012colloquium,fein2019quantum}
and mechanical resonators in optomechanical systems, where light is used to control and measure the mechanical resonator \cite{aspelmeyer2014cavity}. 
It is now possible to prepare nonclassical states of bulk acoustic resonators \cite{chu2018creation,satzinger2018quantum,Luepke2022parity,bild2023schrodinger}, and entanglement between mechanical resonators \cite{riedinger2018remote,Marinkovic2018optomech,ockeloen-korppi2018stabilized}. The most promising platforms for the mentioned tests include levitated systems \cite{millen2020optomechanics,moore2021searching} for which ground state cooling has been demonstrated recently \cite{Delic2020:cooling}. 

Proposed applications in the realm of gravity range from force and acceleration sensing for inertial navigation and geodesy to fundamental research on the quantization of the gravitational field and gravitational wave detection. In particular, mechanical quantum systems can provide information about the time evolution of gravitational and inertial effects depending on the time scales of their dynamics. For example, mechanical resonators can be employed as narrow-band sensors for oscillating gravitational fields, as proposed already in the 1960s by Weber \cite{PhysRevLett.20.1307}. More generally, quantum-enhanced detecting of small displacements of mechanical resonators has been a key goal, ranging from the recent use of squeezed light in  LIGO \cite{jia2024squeezing} to back-action evasion schemes devised for cavity optomechanics \cite{purdy2013observation}. 
While these measurements usually provide sensitivities just below the Standard Quantum Limit, genuine quantum advantage for sensing can be better quantified by the quantum Fisher information (QFI) and the Cramér-Rao bound.  The case of force sensing with optomechanical systems in the nonlinear regime has been analyzed in terms of the QFI in \cite{armata2017quantum,qvarfort2018gravimetry}, including for time-dependent gravitational fields \cite{Qvarfort2021}. 
In the domain of fundamental research, mechanical resonators have also been 
proposed as promising quantum sensors to detect dark matter~\cite{kilian2024dark,carney2021mechanical,carney2021ultralight} and gravitons~\cite{Carney2024:graviton,tobar2024detecting}. It has also been suggested to search for empirical evidence for the quantization of the gravitational field by gravitationally coupling free-falling masses \cite{Bose2017:spin,Marletto2017} or the mechanical elements of two quantum optomechanical systems \cite{Wan2017,miao2020quantum} under the overarching term of gravitationally induced entanglement. The alternative option of classical gravity represented by models like the Schrödinger-Newton equation (SNE) \cite{Diosi1984gravitation,Carlip2008isquantum,Giulini2011classical}, hybrid classical-quantum models \cite{oppenheim2023postquantum,oppenheim_gravitationally_2023}, 
and collapse models \cite{bassi2013models} may be tested by a range of effects on optomechanical resonators such as anomalous heating \cite{carlesso2022present}.

\section{Current and future challenges}

While quantum metrology of gravity with mechanical quantum systems 
has developed strongly on the demand side, it faces considerable challenges on the supply side. In particular, most proposed applications need (i) large mass, (ii) large delocalization of quantum states 
and (iii) long coherence times. Let us expand briefly on these criteria. (i) The need for large masses results from the fact that gravitational forces are very weak and compete with 
electromagnetic forces, but scale with the masses of the probe and source particle. Some proposed applications require quantum systems of large mass in order to source a gravitational field \cite{Carlesso2019:testing,Miao2020:quantum,Plato2023:enhanced}, and signal-to-noise ratios tend to increase with a positive power of the sensor mass \cite{spengler_perspectives_2022}.
(ii) The arguments for large delocalization of the involved quantum states for given mass are similar: On the sensing side, the bounds for the fundamental sensitivities generally scale with the variance of the employed probe state in the degrees of freedom affected by the gravitational signal that is to be probed. On the sourcing side, larger delocalization of the relevant degrees of freedom implies larger quantum effects in the gravitational signal. 
In this context, the metrological perspective on such tests of gravitationally induced entanglement becomes clear when taking into account that a necessary condition for their realizablity is that source and sensor of the gravitational field must both be sensitive to the respective other's gravitational field \cite{Plato2023:enhanced,Pedernales2023on}. (iii) In most cases, the need for large coherence times is simply explained by the advantage of coherently accumulating a signal. Further technical challenges include efficient cooling, keeping heating at bay, and fast and efficient measurement of quadratures. In many cases, mechanical resonators of low frequencies are advantageous, which are however, more difficult to cool. The greatest experimental challenge in precisely measuring the gravitational constant $G$ is a precise determination of the involved masses. 
A promising complement to metrology with the centre-of-mass motion of levitated or free-falling mechanical systems is to employ their orientation degrees of freedom \cite{stickler2021quantum}; rotational dynamics comes with inherent nonlinearities that are absent in the free or harmonic linear motion of a nanoparticle.

In tests of gravitationally induced entanglement, decoherence is the major challenge as it destroys the entanglement that would rule out measurement-feedback models or incoherent theories of gravity \cite{kafri2015bounds,tilloy2016sourcing,oppenheim2023postquantum,oppenheim_gravitationally_2023}. 
A current theoretical challenge lies in formulating  theories of a classical gravitational field that interacts with quantum matter through an effective measurement-feedback process in a relativistically consistent manner \cite{tilloy2024general}.
In contrast, the nonlinear SNE \cite{Diosi1984gravitation,Carlip2008isquantum,Giulini2011classical} describes the gravitational self-interaction of a wave function, leading to 
testable effects on optomechanical resonators 
that differ from those of stochastic models of gravity \cite{Yang2013macroscopic,Grossardt2016optomechanical}. 
Still, the SNE could be falsified in the same way as incoherent gravity theories: by observing gravitational entanglement \cite{gruca2024correlations}.

\begin{table}
    \centering
    \includegraphics[width=15cm,angle=0]{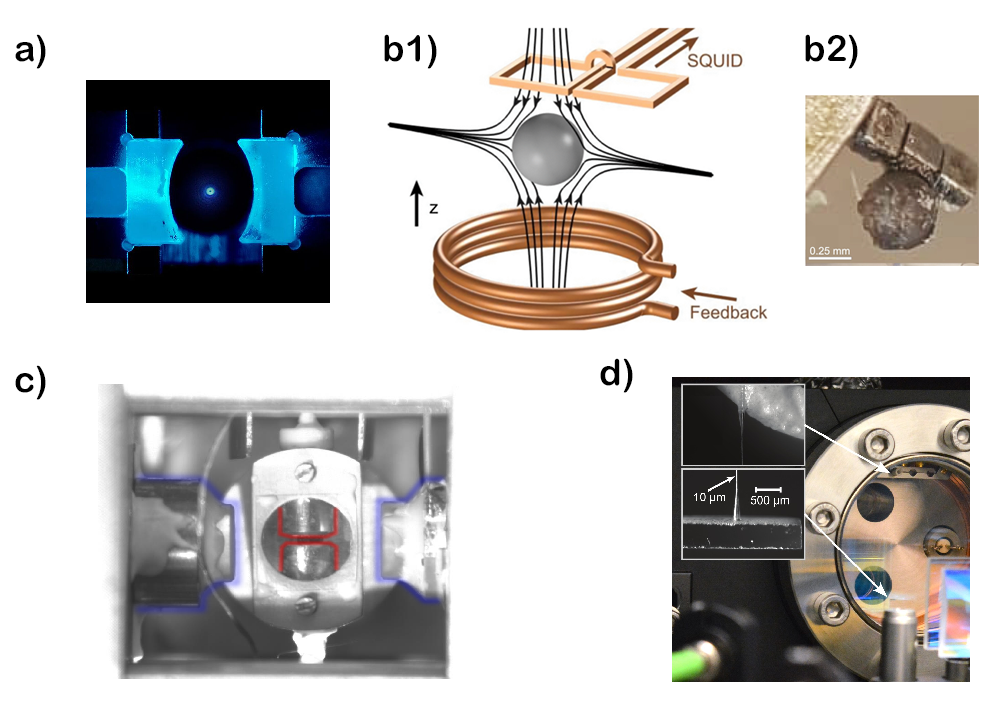}
    \caption{ 
    There is a variety of platforms based on mechanical quantum systems which can be employed for quantum metrology. This figure shows a few examples: a) Levitated nanoparticle (in the center) in optical trap crated within a cavity (light structures left and right) (image courtesy of Iurie Coroli, Kahan Dare, Manuel Reisenbauer, Uros Delic and Lorenzo Magrini, Aspelmeyer Group, University of Vienna, system reported on in \cite{Delic2020:cooling}) b1) and b2) Levitated Superconducting Microspheres (figure in b1 from \cite{Hofer2023:high} and picture in b2 from \cite{Fuchs2024:meas}) c) Levitated particle (in the center between pole pieces highlighted in red, not visible) in hybrid trap composed of a Paul trap (pole pieces highlighted in red) and an optical trap (cavity mirrors highlighted in blue) (image courtesy of James Millen, setup reported on in \cite{Millen2015:cavity})  d) Miniature torsion balance built from a silica slab (lower inset) attached to fiber (image courtesy of Sofia Agafonova, setup reported on in \cite{agafonova2024laser}). 
    }
    \label{fig:systems}
\end{table}

{A common issue in testing, for example, classical or semiclassical theories of gravity, collapse models, and the presence of non-standard weak noise sources due to dark matter is the robust certification of quantum signatures in the mechanical motion. }
As the experiments become significantly more demanding with growing mass, we need robust and data-efficient schemes to certify quantumness of the time evolution and discriminate quantum and classical states instead of resource-hungry state tomography 
and Wigner-function-tomography methods.

\section{Advances in science and technology to meet challenges}

Regarding efficient methods to certify quantumness in the mechanical system, entanglement-breaking models could be probed directly by entanglement witnesses or LOCC process witnesses \cite{Lami2024testing}, adopted and tailored to the measurement settings in (levitated) optomechanics. Other possibilities that promise lower measurement requirements need to be developed further: optimal Wigner negativity witnesses \cite{filip2011detecting,chabaud2021witnessing,zaw2024certifiable}, Tsirelson's inequality \cite{tsirelson2006how,zaw2022detecting,jayachandran2023dynamics,zaw2023dynamics,plavala2024tsirelson}, or methods based on truncated moment sequences \cite{PhysRevA.96.032312,PhysRevA.102.052406}. 

Sensitivities may be improved by combining many detectors into coherent detector networks that are read out collectively or by coupling many detectors to one mode. The maximal scaling of the Cramér-Rao bound that can be achieved that way is the inverse of the number of detectors in the network \cite{fraisse_coherent_2015,braun_coherently_2014}. 
However, noise 
affects this scaling and determines an optimal network size. 

To address the  challenges in precisely determining the masses in experiments that measure $G$, it might be useful to look at mass measurements in other fields. Substantial progress has been achieved in recent years in time-of-flight (TOF) mass analysis in the ultra-high mass range of molecules, meaning up to about $10^6$u $\simeq 10^{-17}$g \cite{lee_limitation_2011,lee_high_2011}. This is still very far from the mg-masses considered for typical trapped particles in the context of gravitational force measurements.  However, the relative precision of mass measurements of about $10^{-5}$ to possibly $10^{-6}$ in this field, suggest that controlled electric charging of trapped nano-particles and measurement of their response to controlled electric or magnetic fields might be a path towards more precise estimations of their mass and hence measurements of $G$. Recently, the simultaneous measurement of the recoil and single electrical charges allowed to detect individual nuclear decays in a trapped nano-particle \cite{wang_mechanical_2024}.

The quantum Cramer–Rao bound, based on QFI, sets the ultimate 
sensitivity limit for parameter estimation.  Quantum metrology protocols optimize the QFI over states, and resources, such as measurement time 
or the number of probes
to boost sensitivity and reduce noise. 
Bayesian inference and hypothesis testing can further refine 
parameter estimates and shorten observation times by enabling robust conclusions from limited data. Optimal quantum hypothesis testing methods, combined with quantum metrology protocols and resources like entangled or squeezed light, can achieve lower error probabilities than classical methods within specific parameter ranges~\cite{marchese2021optomechanical}, with limits set by the Helstrom bound. 
In addition, active error-correction methods can be utilised to improve the coherence time of the detectors \cite{PhysRevLett.112.080801,PhysRevLett.112.150802,PhysRevLett.112.150801}.
Together, these methods will advance quantum sensing by providing sharper estimates and parameter constraints across a wide range of applications, such as detecting gravitational waves, dark matter signatures, and modifications to quantum mechanics. They also bridge theory and experiment by identifying optimal setups, system geometry, resources, and parameter ranges for precise signal detection.

\section{Concluding remarks}

{Mechanical} quantum systems are promising tools for metrology 
of gravity and inertial effects. The 
ability to control such systems on the quantum level, including the preparation of highly non-classical probe states, may be used to improve sensitivities and 
potentially access quantum properties of the gravitational field. To this end, the main challenges 
include 
increasing the mass, improving cooling, and keeping decoherence and heating at bay by isolating the massive quantum systems from gas collisions (ultra-high vacuum),
thermal reservoirs (cryostats), and mechanical vibrations. These challenges are 
currently being addressed by experimental groups around the world.
On the theory side, further novel avenues for testing gravitational effects with 
{mechanical} systems in the quantum regime are needed, in addition to schemes to mitigate decoherence. 

\section*{Acknowledgements}

S.Q.~is funded in part by the Wallenberg Initiative on Networks and Quantum Information (WINQ) and in part by the Marie Skłodowska--Curie Action IF programme \textit{Nonlinear optomechanics for verification, utility, and sensing} (NOVUS) -- Grant-Number 101027183. Nordita is funded in part by NordForsk. DB and HU acknowledge support by the EU EIC Pathfinder project QuCoM (101046973). HU thanks for support the UKRI EPSRC (EP/W007444/1, EP/V000624/1 and EP/X009491/1), the Leverhulme Trust (RPG-2022-57) and the QuantERA II Programme (project LEMAQUME) that has received funding from the European Union’s Horizon 2020 research and innovation programme under Grant Agreement No 101017733. DR acknowledges support by the Deutsche Forschungsgemeinschaft (DFG, German Research Foundation) under Germany’s Excellence Strategy – EXC-2123 QuantumFrontiers – 390837967.
MMM acknowledges support from the Walter Benjamin Programme (project number 510053905).

\bibliography{refs}

\end{document}